\documentclass[twocolumn,superscriptaddress,prl,amsmath,amstex,amssymb,citeautoscript]{revtex4-1}
\pdfoutput=1
\usepackage[english]{babel}
\usepackage{letltxmacro}
\usepackage{latexsym}
\usepackage[utf8]{inputenc}
\LetLtxMacro{\ORIGselectlanguage}{\selectlanguage}
\makeatletter
\DeclareRobustCommand{\selectlanguage}[1]{%
  \@ifundefined{alias@\string#1}
    {\ORIGselectlanguage{#1}}
    {\begingroup\edef\x{\endgroup
       \noexpand\ORIGselectlanguage{\@nameuse{alias@#1}}}\x}%
}
\newcommand{\definelanguagealias}[2]{%
  \@namedef{alias@#1}{#2}%
}
\makeatother
\definelanguagealias{en}{english}
\definelanguagealias{English}{english}
\usepackage{graphicx}
\usepackage{amsmath}
\usepackage{amsfonts}
\usepackage{amsthm}   %
\usepackage{amssymb}
\usepackage{bm}
\usepackage{bbm}
\usepackage{color}
\usepackage[percent]{overpic}
\usepackage{soul} 
\usepackage{amssymb}
\usepackage{wasysym}
\usepackage{dsfont}
\usepackage{float}
\usepackage{mathtools}

\usepackage{hyperref}
\hypersetup{
    unicode=false,          %
    pdftoolbar=false,        %
    pdfmenubar=true,        %
    pdffitwindow=false,     %
    pdfstartview={FitH},    %
    pdftitle={},    %
    pdfauthor={Authors},     %
    pdfsubject={},   %
    pdfcreator={},   %
    pdfproducer={}, %
    pdfkeywords={}, %
    pdfnewwindow=true,      %
    colorlinks=true,       %
    linkcolor=black,          %
    citecolor=blue,        %
    filecolor=magenta,      %
    urlcolor=black           %
}

\newcommand{\be}{\begin{equation}}
\newcommand{\ee}{\end{equation}}
\newcommand{\bea}{\begin{eqnarray}}
\newcommand{\eea}{\end{eqnarray}}

\usepackage{graphicx}
\usepackage{verbatim}

\newcommand{\ket}[1]{\mbox{$| #1 \rangle$}}
\newcommand{\kket}[1]{\mbox{$| #1 \rangle\!\rangle$}}

\newcommand{\abs}[1]{\lvert#1\rvert}

\newcommand{\tr}{\mathrm{tr}}

\usepackage{soul}

\newtheorem*{lemma*}{Lemma}

\begin{document}

\title{Finite-time teleportation phase transition in random quantum circuits}

\author{Yimu Bao}
\affiliation{Department of Physics, University of California, Berkeley, California 94720, USA}

\author{Maxwell Block}
\affiliation{Department of Physics, University of California, Berkeley, California 94720, USA}

\author{Ehud Altman}
\affiliation{Department of Physics, University of California, Berkeley, California 94720, USA}
\affiliation{Materials Sciences Division, Lawrence Berkeley National Laboratory, Berkeley, California 94720, USA}

\begin{abstract}
How long does it take to entangle two distant qubits in a quantum circuit evolved by generic unitary dynamics? We show that if the time evolution is followed by measurements of all but two {\em infinitely} separated test qubits, then the entanglement between them can undergo a phase transition and become nonzero at a finite critical time $t_c$. The fidelity of teleporting a quantum state from an input qubit to an infinitely distant output qubit shows the same critical onset.
Specifically, these finite-time transitions occur in short-range interacting two-dimensional random unitary circuits and in sufficiently long-range interacting one-dimensional circuits.
The phase transition is understood by mapping the random continuous-time evolution to a finite-temperature thermal state of an effective spin Hamiltonian, where the inverse temperature equals the evolution time in the circuit. 
In this framework, the entanglement between two distant qubits at times $t>t_c$ corresponds to the emergence of long-range ferromagnetic spin correlations below the critical temperature. 
We verify these predictions using numerical simulation of Clifford circuits and propose potential realizations in existing platforms for quantum simulation.
\end{abstract}

\maketitle

The dynamics of entanglement in many-body quantum systems is the focus of intense theoretical~\cite{kim2013ballistic,nahum2017quantum,SkinnerNahum2018measure,LiFisher2018ZenoEffect} and experimental interest~\cite{kaufman2016quantum,arute2019quantum,choi2023preparing,noel2022measurement}; indeed, it provides crucial insights for understanding the capacity of physical systems to process quantum information as well as the computational complexity involved in simulating their dynamics~\cite{vidal2003efficient}.
In generic unitary evolution with short-range interactions, the Lieb-Robinson bound~\cite{lieb1972finite} ensures that quantum entanglement propagates along light cones. Thus, two degrees of freedom separated by a distance $L$ take a time of order $L$ to get entangled. 

Entanglement can be created much faster by supplementing unitary evolution with measurements~\cite{briegel1998quantum,li2021conformal,cotler2023emergent}. 
As a simple example, we consider a chain of qubits initialized in a product of Bell pairs on the \emph{odd} links, which can be prepared from a product state by a single layer of two-qubit gates.
By performing Bell measurements on the {\em even} links, one can create a Bell pair of the (unmeasured) first and last qubit.
As another example, a two-dimensional cluster state can be used for measurement-based quantum computation~\cite{raussendorf2001one}; one can create any desired entangled state by appropriate local measurements. 

In the schemes described above, entanglement is generated over arbitrarily long distances using a unitary circuit of constant depth followed by a single layer of measurements. 
These states are said to possess finite localizable entanglement for two distant qubits (hence infinite entanglement length)~\cite{verstraete2004entanglement,popp2005localizable}.
However, the examples above are highly fine-tuned. 
It is natural to ask how long it would take to create quantum correlations between distant qubits using generic unitary evolution followed by local measurements.

\begin{figure}[t!]
\centering
\includegraphics[width=0.47\textwidth]{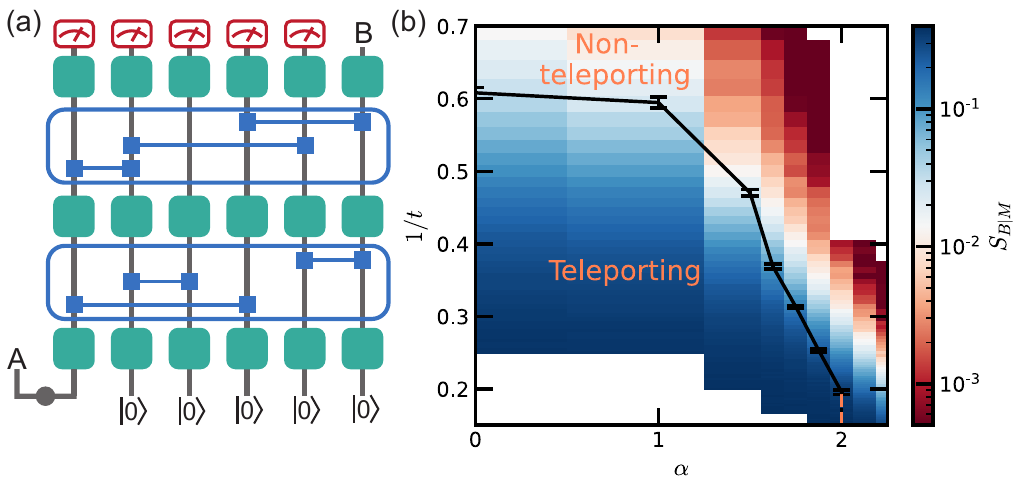}
\caption{(a) Random quantum circuit on $N$ qubits. In each time step $\delta t$, we apply single-qubit Haar random gates to every qubit followed by $N\delta t$ two-qubit Haar random gates. The distribution of two-qubit gates is determined by the geometry of the circuit. Initially, the first qubit is maximally entangled with a reference qubit $A$, while the remaining qubits are prepared in $\ket{0}$. We consider the entropy of an output qubit $B$ conditioned on local measurements on the rest of the qubits. (b) Phase diagram of one-dimensional long-range unitary circuits with power-law decaying interaction. The black markers represent the inverse critical time $1/t_c$ as a function of power-law exponent $\alpha$. The transition exists for $\alpha \leq 2$, indicated by the pink dashed line. The color indicates the conditional entropy $S_{B|M}$.}
\label{fig:setup}
\end{figure}

In this Letter, we show that the creation of states with infinite entanglement length can occur as a phase transition at a critical time of order one. %
In the simplest setup, an initial product state is evolved for a time $t$, after which all but two infinitely separated qubits are measured. 
In two (or higher) dimensional systems with short-range interactions, entanglement between the two distant qubits onsets at a critical time $t_c$.
The same is true for one-dimensional systems with sufficiently long-range interactions.
An equivalent scheme, shown in Fig.~\ref{fig:setup}(a), is to consider the teleportation from an input qubit to an infinitely distant output qubit after measuring all other output qubits, which can be achieved with nonvanishing fidelity after time $t_c$.

We provide a theoretical picture of this transition by mapping the random circuit evolution to an effective equilibrium problem. 
This approach builds on recent developments in describing entanglement dynamics through mapping circuits consisting of random unitaries to the statistical mechanics of classical spins located at the space-time positions where the gates operate~\cite{hayden2016holographic,nahum2018operator, vasseur2019entanglement,bao2020theory,jian2020measurement}. 
In the case of continuous time evolution, the classical spin model can be viewed as imaginary time evolution generated by an effective quantum Hamiltonian~\cite{bao2021symmetry,block2022measurement}.

Previously, this mapping was primarily applied to understand the steady-state entanglement properties, which are determined by the ground state of the effective Hamiltonian (i.e. infinite imaginary time evolution).
Similarly, the finite-time evolution, which we consider here, is related to a thermal state of the effective Hamiltonian (i.e. finite imaginary time evolution).
The key point is that a finite-temperature transition in the thermal state indicates a finite real-time transition in the circuit.

We demonstrate this phenomenon by considering the transition in continuous-time random unitary circuits (RUC) with different architectures including: two-dimensional short-range systems, all-to-all coupled systems, and one-dimensional long-range systems with power-law decaying interactions.
In the last case, our theory predicts a finite-time transition for power-law exponent $\alpha \leq 2$, and specifically a Kosterlitz-Thouless (KT) like transition at $\alpha = 2$. 
We corroborate these predictions with numerical simulations of random Clifford circuits.

Before proceeding, we remark on a related paper by Napp et al.~\cite{napp2022efficient} dealing with the sampling complexity of shallow two-dimensional brick-layer RUCs. This work claimed and provided evidence that approximate sampling from such circuits is hard if the depth $t$ is above a threshold $t_c$ of $O(1)$ while it is easy for $t < t_c$. Sampling requires measuring all qubits following the final layer of unitary gates. The essence of the argument is that this network can be contracted sideways, showing it is equivalent to simulating the dynamics of a one-dimensional quantum circuit with measurements.
Hence, one expects a phase transition in sampling complexity in shallow two-dimensional RUCs which is of the same universality as the measurement-induced transition in one dimension~\cite{LiFisher2018ZenoEffect,SkinnerNahum2018measure,li2019measurement,choi2020quantum,gullans2020dynamical}.
In our discussion, we argue heuristically that this sampling transition can be understood as a specific example of the teleportation transition, and may therefore occur in a broad class of systems for which the effective Hamiltonian exhibits a finite-temperature transition.

\emph{Setup and theoretical framework.}---
Our model consists of $N$ qubits with $N-1$ qubits initialized in a product state and a single qubit prepared in a maximally entangled state with the reference $A$.
In each time step $\delta t$, we apply a layer of single-qubit Haar random unitary gates followed by $N\delta t$ two-qubit Haar random unitary gates [Fig.~\ref{fig:setup}(a)]. 
The sites $(i,j)$ on which each two-qubit gate operates are drawn independently from a distribution $P(i,j)$, which depends on the specific models we discuss below.
The single-qubit gates do not generate entanglement and are introduced only for analytical convenience~\footnote{We note that single-qubit gate on site $i$ can in principle change the circuit ensemble provided that no two-qubit Haar random gate acts on $i$.}.
After evolving for time $t$, we measure all $N-1$ qubits except for a distant qubit $B$.

The resulting fidelity of teleportation between qubits $A$ and $B$ can be quantified (without considering an explicit decoding scheme) by the entanglement entropy of $B$, conditioned on the measurement outcomes of qubits $M$~\cite{SOM}.
To analytically determine the conditional entropy $S_{B|M}$ averaged over circuit realizations and measurement outcomes, we formulate it as the $n \to 1$ limit (replica limit) of the quantities~\cite{bao2020theory}
\begin{equation}
S_{B|M}^{(n)} = \frac{1}{1-n}\log\frac{\overline{\sum_m \tr\tilde{\rho}_{B,m}^n}}{\overline{\sum_m \tr\tilde{\rho}_{m}^n}}, \label{eq:SBM}
\end{equation}
where $\tilde{\rho}_m := \hat{P}_m \rho \hat{P}_m$ is the projection of the density matrix onto the set of measurement outcomes labeled by $m$, and the overline indicates the average over the Haar ensemble.
Accordingly, the probability for this set of measurement outcomes is $p_m=\tr(\rho \hat{P}_m)=\tr\tilde{\rho}_m$, and the normalized density matrix is $\rho_m= \tilde{\rho}_m/p_m$.
We note that the $S^{(n)}_{B|M}$ are not precisely the conditional R\'enyi entropies because the average is taken inside the logarithm.

The simplest quantity that captures the qualitative features of $S_{B|M}$ is $S_{B|M}^{(2)}$, although the critical exponents of their respective transitions may be different~\cite{vasseur2019entanglement,bao2020theory,jian2020measurement}.
The quantity $S^{(2)}_{B|M}$ involves the second moments of the density matrix, which can be determined from the double density matrix $\rho \otimes \rho$.
Formally, $\rho \otimes \rho$ can be represented as a state vector $\kket{\rho}$ in the replicated Hilbert space $\mathcal{H}^{(2)} := (\mathcal{H} \otimes \mathcal{H}^*)^{\otimes 2}$, where $\mathcal{H}$ ($\mathcal{H}^*$) denotes the ket (bra) Hilbert space.
A unitary gate $U$ in the circuit acts as $\mathcal{U} = (U \otimes U^*)^{\otimes 2}$ on $\kket{\rho}$.
Hence, the replicated density matrix undergoes unitary evolution $\kket{\rho(t)} = \prod_{\tau = 1}^{N_t} \mathcal{U}_{2,\tau} \mathcal{U}_{1,\tau}\kket{\rho(0)}$, where $\mathcal{U}_{1,\tau}$ and $\mathcal{U}_{2,\tau}$ denote the layer of single- and two-qubit gates in each time step $\tau$, respectively.

The average dynamics of the double density matrix can be analytically mapped to imaginary time evolution under an effective Hamiltonian~\cite{bao2021symmetry,block2022measurement}.
First, the average over single-qubit gates effects a projection from a sixteen-dimensional local Hilbert space to the two-dimensional Hilbert space of a spin-$1/2$~\cite{SOM}.
Then, the layer of two-qubit gates reduces to a transfer matrix for the transition amplitude between the spin-$1/2$ configurations in consecutive time steps, $\mathcal{T} = \mathds{1} + N\delta t \sum_{ij} P(i,j) \overline{\mathcal{U}_{2,\tau}(i,j)}$. 

\begin{figure*}[t!]
\centering
\includegraphics[width=\textwidth]{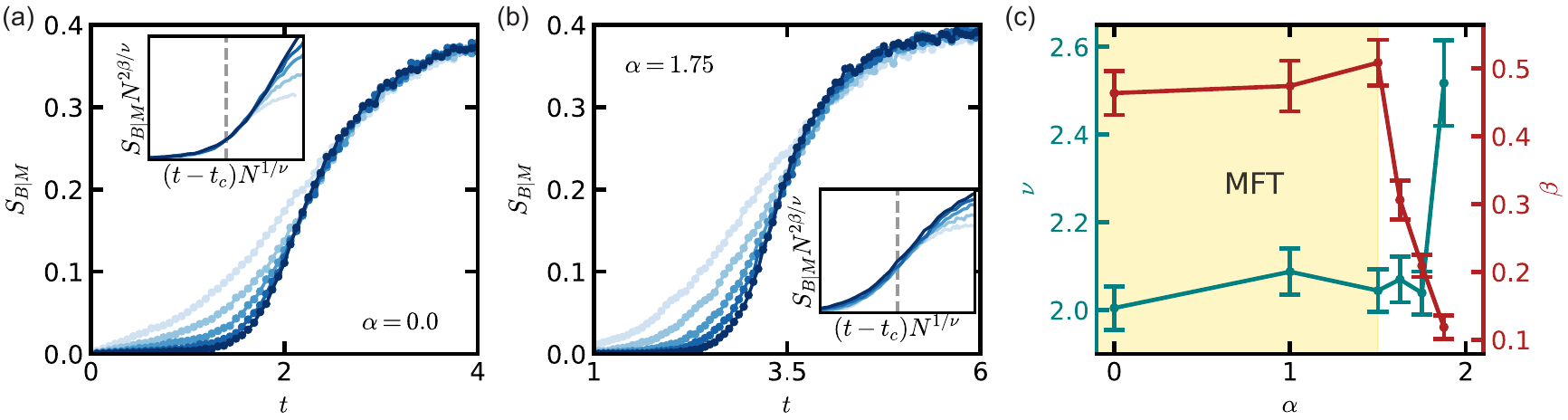}
\caption{Finite-time transition in one-dimensional long-range interacting random circuits. (a,b) The conditional entropy $S_{B|M}$ in circuits with power-law exponents $\alpha = 0$ [all-to-all, panel (a)] and $\alpha = 1.75$ [panel (b)] plotted as a function of time $t$ for various system sizes $N$ from $32$ to $512$ indicated by increasing opacity. (inset) Finite-size scaling collapse using Eq.~\eqref{eqn:scaling-form}. The grey dotted line indicates $t_c$. For the all-to-all circuit ($\alpha = 0$), we obtain critical exponents $\nu \approx 2.0 $, $\beta \approx 0.46$, and critical time $t_c \approx 1.6$. For $\alpha = 1.75$, we obtain $\nu \approx 2.0 $, $\beta \approx 0.20$, and critical time $t_c \approx 2.1$. (c) Critical exponents $\nu$ and $\beta$ for $\alpha < 2$. The exponents for $\alpha \leq 1.5$ agree with the prediction from mean-field theory. Moreover, near $\alpha=2$, $\nu$ begins to diverge, as expected near a KT-like transition. The finite-time transition does not exist for $\alpha > 2$. The numerical results are averaged over $1.5\cdot10^4$ random circuit realizations.}
\label{fig:long-range_ckt}
\end{figure*}

The transfer matrix $\mathcal{T}$ can be viewed as the infinitesimal imaginary time evolution generated by an effective quantum Hamiltonian operating on spin-$1/2$ degrees of freedom,
$\mathcal{T} = e^{-\delta t H_{\text{eff}}}$. 
For our circuit, the effective Hamiltonian takes the form
\begin{align}
     H_{\text{eff}} &= \sum_{i,j}J_{ij} \left[ -\frac{2}{5}\sigma_i^z\sigma_j^z - \frac{1}{10}\sigma_i^y \sigma_j^y - \frac{1}{5}(\sigma_i^x + \sigma_j^x) \right],
     \label{eq:Heff}
\end{align}
where the coupling $J_{ij} = N P(i,j)$ is given by the average number of two-qubit gates acting between qubit $i$ and $j$ in every unit time~\footnote{We note that because only $N$ gates are applied per unit time, $\sum_{(i,j)} J_{ij} = N$, and therefore the energy of the Ising model is always extensive.}.
Accordingly, the replicated, un-normlized, density matrix evolves as $\kket{\rho(t)} = e^{-tH_{\text{eff}}}\kket{\rho(0)}$.
We note that the Hamiltonian exhibits a global Ising symmetry generated by $\prod_i\sigma^x_i$, which stems from the invariance of $\mathcal{U}$ under the permutation of two copies of ket (or bra) Hilbert space.

The effective imaginary time evolution above yields a thermal state of the ferromagnetic Ising Hamiltonian in Eq.~\eqref{eq:Heff} at inverse temperature $t$.
For two-dimensional RUCs, the associated two-dimensional Ising model will undergo a ferromagnetic transition at a temperature corresponding to a finite critical time $t_c$.
Similarly, for one-dimensional RUCs, the associated one-dimensional Ising model can exhibit a finite-temperature transition provided the unitary couplings decay with a sufficiently small power of distance $\alpha\leq 2$~\cite{ruelle1968statistical,dyson1969existence,thouless1969long,anderson1971some}.
This finite-temperature phase transition implies a transition in the output state of the circuit occurring at a finite time.

This transition manifests in the conditional entropy $S^{(2)}_{B|M}$.
The projective measurements of the output state play a crucial role in revealing the transition: they impose a boundary condition in the finite imaginary time evolution that preserves the Ising symmetry~\cite{bao2020theory,jian2020measurement}.
In more detail, $S^{(2)}_{B|M}$ is mapped to the excess free energy associated with imposing symmetry-breaking fields only at the space-time locations of qubit $A$ and $B$.
This can be further reduced to the imaginary time order parameter correlation function, $S^{(2)}_{B|M} \sim \langle \sigma_B^z(t) \sigma_A^z(0)\rangle$~\cite{SOM}.
Consequently, $S^{(2)}_{B|M}$ is non-decaying in the ordered phase ($t > t_c$) due to the long-range order in the Ising model, whereas $S^{(2)}_{B|M}$ rapidly decays to zero in the disordered phase ($t < t_c$).

We remark that there is no finite-time transition in the purity or entanglement entropy of an extensive subsystem in the output state. In the effective spin model, such quantities involve a symmetry breaking field at the final time ~\cite{hayden2016holographic,nahum2018operator,vasseur2019entanglement}.
Since we are concerned with spontaneous symmetry breaking in the slab, whose thickness is the evolution time of order one, the symmetry breaking fields necessarily eliminate the transition.

\emph{Examples and numerical results}.---
We demonstrate the finite-time teleportation transition predicted above in three exemplary models: (1) all-to-all interacting quantum circuits, (2) one-dimensional quantum circuits with power-law decaying long-range interactions, and (3) two-dimensional quantum circuits with short-range interactions. 
To verify our theoretical predictions, we compute the conditional entropy in random Clifford circuits, which can be efficiently simulated~\cite{gottesman1998heisenberg,aaronson2004improved}.
Although Clifford circuits, which only form a unitary 3-design~\cite{webb2015clifford}, are not the same as Haar random circuits, they still exhibit a finite-time transition with the same qualitative behavior.

First, we consider the circuit with all-to-all unitary gates. 
Within a time step $\delta t$, each two-qubit gate is drawn independently and operates on a random pair of qubits $(i,j)$ with equal probability.
Hence, the effective quantum Hamiltonian that describes $S^{(2)}_{B|M}$ has all-to-all couplings $J_{ij} \sim 1/N$~\cite{SOM}.
In the limit $N\to\infty$, the Ising phase transition in this Hamiltonian is described exactly by mean-field theory, which predicts critical exponents $\nu_{MF}=2, \beta_{MF}=0.5$, and a critical time $t_c^{(2)} = 2.0$~\cite{SOM}.
We note that the mean-field theory does not yield a reliable $t_c$ for $S_{B|M}$ as the effective Hamiltonian is derived for an approximate quantity $S^{(2)}_{B|M}$.

To characterize the transition of the conditional entropy $S_{B|M}$, we simulate this quantity in all-to-all Clifford circuits of system sizes up to $N = 512$ as shown in Fig.~\ref{fig:long-range_ckt}(a)~\footnote{We ignore the single-qubit gates in the Clifford simulation as they do not affect the information dynamics.}~\footnote{We note that $S_{B|M}$ saturates to a maximum value $0.4$ in the long time limit, which is universal for random Clifford evolution~\cite{cotler2023emergent}.}.
We perform a finite-size scaling analysis based on the scaling formula for order parameter correlation function to extract critical exponents~\cite{SOM}:
\begin{align}
S_{B|M}(t,N) = N^{-2\beta/\nu} \mathcal{F} \big( (t - t_c) N^{1/\nu}\big).
\label{eqn:scaling-form}
\end{align}
This analysis yields critical exponents $\nu=2.1 \pm 0.2, \beta=0.4 \pm 0.1$, which are in close agreement with the predictions of the mean-field theory, and also the critical time $t_c \approx 1.6$.

Next, we consider a one-dimensional array of $N$ qubits evolving with power-law decaying couplings and periodic boundary conditions.
Here, for each two-qubit gate, we independently choose a random pair of sites $(i,j)$ with a probability $P(i,j) \propto 1/\abs{i-j}^\alpha$.
The effective model for this circuit is a one-dimensional finite-width classical Ising model with long-range coupling $J_{ij} \sim 1/|i - j|^\alpha$.

This model is in the same universality class as the one-dimensional long-range classical Ising chain at finite temperature, which has been extensively studied and shown to have an ordering transition when $\alpha \leq 2$~\cite{ruelle1968statistical,dyson1969existence,thouless1969long,anderson1971some}, with KT universality at $\alpha=2$~\cite{kosterlitz1976phase,cardy1981one,bhattacharjee1981some,bhattacharjee1982n,imbrie1988intermediate, luijten2001criticality}.
Furthermore, for $3/2< \alpha < 2$, the transition features continuously varying critical exponents, whereas for $\alpha \leq 3/2$, it is described by mean-field theory with $\alpha$-independent exponents~\cite{kosterlitz1976phase}.

These predictions from the classical Ising chain are borne out clearly in our Clifford numerics.
For $\alpha \leq 2$, we simulate $S_{B|M}$ for $A$ and $B$ separated by $N/2$ sites and observe a crossing for various $N$, as exemplified at $\alpha = 1.75$ in Fig.~\ref{fig:long-range_ckt}(b)~\cite{SOM}.
We perform the finite-size scaling to determine the exponents [summarized in Fig.~\ref{fig:long-range_ckt}(c)]~\footnote{
For $\alpha \leq 3/2$ we find approximately constant critical exponents, consistent with mean-field theory. 
In contrast, for $3/2 < \alpha < 2$ we obtain continuously varying critical exponents.
}.
On the other hand, we do not observe a finite-time transition in $S_{B|M}$ for $\alpha > 2$~\cite{SOM}.
A phase diagram is presented in Fig.~\ref{fig:setup}(b).

The point $\alpha=2$ requires special attention. 
Here, the effective model exhibits a finite-temperature KT transition, which does not admit single-parameter scaling as postulated in Eq.~\eqref{eqn:scaling-form}. 
The exponential divergence of the correlation length can be viewed as having $\nu\to\infty$. 
Indeed Fig.~\ref{fig:long-range_ckt}(c) shows a sharp increase of $\nu$ upon approaching $\alpha=2$.
At $\alpha=2$ we compare the observed scaling of $S_{B|M}(t, N)$ to the scaling form $a\exp[1/(\log N+b)]$ expected in an Ising chain with inverse square interaction~\cite{bhattacharjee1981some}. 
We find an accurate fit at the critical time $t_c \approx 5.0$~\cite{SOM}, which supports a KT-like transition. 
However, simulations on larger system sizes are needed to precisely determine the universality of this transition. 

\begin{figure}[t!]
\centering
\includegraphics[width=0.48\textwidth]{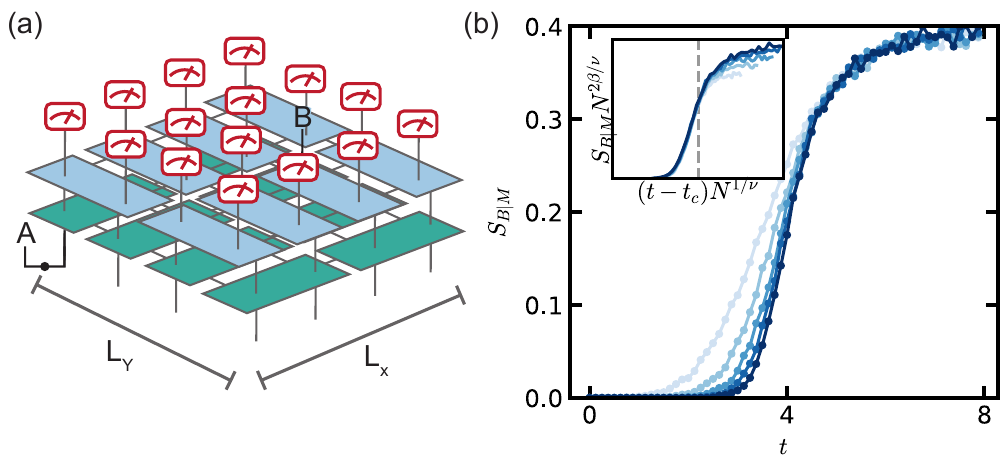}
\caption{Finite-time teleportation transition in two-dimensional short-range random circuits. (a) Schematic of a circuit of size $L_x = L_y = L$. We use periodic boundary conditions and consider reference $A$ to be entangled with an input qubit separated from output qubit $B$ by $L/2$ in both directions. (b) The conditional entropy $I_{B|M}$ plotted as a function of time $t$ for $L$ from $8$ to $24$ indicated by increasing opacity. (inset) Finite-size scaling collapse using Eq.~\eqref{eqn:scaling-form}. We obtain $\nu \approx 1.2 \pm 0.1$, $\beta \approx 0.11 \pm 0.03$, and critical time $t_c \approx 4.2$ (indicated by the grey dashed line). The numerical results are averaged over $9000$ random circuit realizations.}
\label{fig:2d_ckt}
\end{figure}

Last, we consider the finite-time transition in short-range interacting circuits in higher dimensions ($d \geq 2$).
Specifically, we consider $P(i,j)$ to be uniformly distributed over pairs of nearest-neighbor qubits.
The critical exponents extracted from the two-dimensional Clifford simulation are $\nu \approx 1.2$, $\beta \approx 0.11$ [Fig.~\ref{fig:2d_ckt}(b)]. 
These exponents agree with those found in the $(1+1)$D measurement-induced entanglement transition~\cite{zabalo2020critical}~\footnote{The anomalous dimension in two dimensions is given by $\eta = 2\beta/\nu \approx 0.2$ in agreement with the result of~\cite{zabalo2020critical}.}.
This result is indeed expected by mapping the dynamics of a finite depth two-dimensional brick-layer RUC with final-time measurements to monitored quantum dynamics in one dimension~\cite{napp2022efficient}.

\emph{Discussion.}---
The above analysis of two-dimensional circuits suggests that the finite-time teleportation transition may generally correspond to a transition in approximate sampling complexity~\cite{napp2022efficient,maskara2022complexity}.
Specifically, we consider the problem of sampling measurement outcomes from $N$ qubits initialized in a product state and evolved under a finite-time RUC.
To draw a connection to the teleportation transition, we divide the output qubits into three regions: $\mathcal{A}$ and $\mathcal{B}$, each with a sub-extensive number of qubits $N^\gamma$ with $0<\gamma<1$, and $M$, the remaining qubits.

In the teleporting phase ($t > t_c$), measurements on $M$ generate long-range entanglement between subsystems $\mathcal{A}$ and $\mathcal{B}$. 
In the spin model, $S_{\mathcal{B}|M}$ is the excess free energy of imposing a domain wall separating $\mathcal{A}$ from $\mathcal{B}$, which scales as a power law of $\min(\abs{\mathcal{A}}, \abs{\mathcal{B}})$ in the ordered phase~\footnote{The sub-extensive symmetry breaking boundary conditions on $\mathcal{A}$, $\mathcal{B}$ do not destroy the long-range order in the effective spin model.}.
Thus, we expect approximate sampling from the pure joint state $\ket{\psi_\mathcal{A} \psi_\mathcal{B}}$ to be as complex as sampling from a Haar-random state of a sub-extensive power-law number of qubits, which is believed to be classically hard~\cite{bouland2019complexity}.

On the other hand, in the non-teleporting phase ($t < t_c$), the effective model has a finite correlation length $\xi$, i.e. sampling from a given qubit is independent from sufficiently distant qubits.
Indeed, it has been shown for brick-layer circuits that approximate sampling can be achieved by patching simulations of subregions of size $O[(\log N)^d]$ together, resulting in a $\text{Poly}(N)$ runtime in two dimensions and quasi-Poly$(N)$ runtime in higher dimensions~\cite{napp2022efficient}.
However, establishing a rigorous connection between finite-time teleportation in Haar-random circuits with arbitrary connectivity and sampling complexity remains an open question for future work.

Although 1D short-range RUCs do not feature a finite-time transition, the spin model mapping indicates an exponentially diverging correlation length $\xi \sim \exp(Jt) $ with circuit depth $t$.
This results from the correlation length $\xi \sim \exp(J/T)$ in the 1D quantum Ising model at temperature $T$ with coupling $J$.
Therefore, one can teleport qubits over a distance $N$ in circuits of depth $t \sim \log N$~\footnote{However, the approximate sampling from $\log N$-depth 1D RUCs is tractable since $S_{B|M} = O(\log N)$~\cite{osborne2006efficient}.}.

The teleportation transition we describe can potentially be realized on leading quantum simulation platforms, such as trapped-ion systems, which feature tunable long-range interactions~\cite{porras2004effective}, and two-dimensional superconducting circuits~\cite{arute2019quantum,satzinger2021realizing}\footnote{We remark our protocol demonstrates a distinct teleportation mechanism from those inspired by quantum gravity~\cite{hayden2007black,landsman2019verified,brown2023quantum,schuster2022many}, in which the teleportation distance is limited by the lightcone of the evolution.}.
We note, however, that obtaining the conditional entropy in experiments is challenging as naive evaluation of $S_{B|M}$ requires post-selection on an extensive number of qubits.
Alternatively, one can verify the entanglement by decoding from the output qubit, which is a topic of ongoing research for generic evolution beyond Clifford circuits~\cite{gullans2020scalable,noel2022measurement}.

Our framework is also applicable to studying finite-time transitions in other circuit ensembles.
In circuits with conserved quantities, the effective Hamiltonian is governed by an enlarged symmetry allowing a richer phase structure at finite times~\cite{bao2021symmetry}.
For example, in free fermion dynamics that conserve fermion parity, the effective Hamiltonian exhibits a continuous $U(1)$ symmetry.
In two dimensions, the effective model undergoes a finite-time KT transition and can support power-law decaying $S_{B|M}$, while in dimension $d \geq 3$, the continuous symmetry can be broken, leading to non-decaying $S_{B|M}$.
Moreover, we note that the key dynamical feature that enables the teleportation transition is the protection of quantum information against local measurements.
Thus, we conjecture that the transition can also occur in non-random chaotic Hamiltonian dynamics in which local scrambling protects information.

\begin{acknowledgments}
\emph{Acknowledgements}.---We thank Soonwon Choi, Abhinav Deshpande, Michael Gullans, Nishad Maskara, Norman Yao, and Xiaoliang Qi for useful discussions.
We are grateful to Alex Dalzell for insightful feedback and carefully explaining the work of Ref.~\cite{napp2022efficient}.
This material is based upon work supported by the U.S. Department of Energy, Office of Science, National Quantum Information Science Research Centers, Quantum Systems Accelerator. This work was supported in part by the NSF QLCI program through grant no. OMA-2016245. M. B. acknowledges support through the Department of Defense (DOD) through
the National Defense Science and Engineering Graduate
(NDSEG) Fellowship Program. E. A. acknowledges support from the Gyorgy Chair in Physics at UC Berkeley.
\end{acknowledgments}

\bibliography{refs}

\end{document}


\title{Supplementary Online Material for ``Finite-time teleportation phase transition in random quantum circuits''}
\date{\today}

\author{Yimu Bao}
\affiliation{Department of Physics, University of California, Berkeley, California 94720, USA}

\author{Maxwell Block}
\affiliation{Department of Physics, University of California, Berkeley, California 94720, USA}

\author{Ehud Altman}
\affiliation{Department of Physics, University of California, Berkeley, California 94720, USA}
\affiliation{Materials Sciences Division, Lawrence Berkeley National Laboratory, Berkeley, California 94720, USA}

\maketitle

\tableofcontents

\section{Teleportation in the finite-time circuit}

In the finite-time circuit [shown in Fig.~1(a) of the main text], we are interested in the teleportation from the reference qubit $A$ to the output qubit $B$.
The teleportation is equivalent to transmitting quantum information through a quantum channel from the leftmost qubit in the initial state to the output of the circuit.
The maximum number of teleported qubits (or in our case the fidelity of teleporting a single qubit $A$) is characterized by the channel capacity.
It is worth noting that we here consider the information retained in both the quantum state of qubit $B$ and the classical measurement results in $M$.
Thus, the output of the channel is the entire system including $B$ and $M$.

The channel capacity is given by the maximum amount of coherent information that can be transmitted~\cite{PreskillNotes}.
In Ref.~\cite{choi2020quantum}, the authors showed that the coherent information of such a monitored circuit is given by the entropy of the output $B$ conditioned on measured qubits $M$, which is also the averaged entropy of $B$ over measurement outcomes in $M$, i.e. $\sum_m p_m S_{B,m}$.
We use this quantity throughout the paper to detect the potential transition in teleportation fidelity.

\section{Effective quantum Hamiltonian}
In this section, we show that the finite-time random unitary circuit evolution maps to the thermal state of an effective ferromagnetic Ising Hamiltonian at a finite temperature.
As emphasized in the main text, the finite-temperature ferromagnetic transition in the effective Hamiltonian manifests as the finite-time transition in the maximum teleportation fidelity, i.e. the conditional entropy, $S_{B|M}$.
In particular, we derive the boundary conditions associated with $S_{B|M}$ and show it maps to the order parameter correlation function that detects the transition.

We consider random unitary circuits in Fig.~1(a) of the main text and compute the entropy of the output qubit $B$ conditioned on measurement results on the rest of the output qubits $M$. 
The conditional entropy $S_{B|M}$ provides an upper bound on the teleportation fidelity.
We seek its average value over the ensemble of trajectories defined by random circuit realizations and measurement results on $M$,
\begin{align}
    S_{B|M} = \overline{\sum_m p_m S_{B,m}},
\end{align}
where $\overline{\cdot}$ represents the average over unitary gates, and $m$ labels the measurement results.
Using the framework developed in Ref.~\cite{bao2021symmetry,block2022measurement}, we can express $S_{B|M}$ using the replica method as the $n \to 1$ limit of the replicated quantities
\begin{align}
    S^{(n)}_{B|M} = \frac{1}{1-n}\log \left( \frac{\overline{\sum_m \tr \tilde{\rho}_{B,m}^n}}{\overline{\sum_m \tr \tilde{\rho}_{m}^n}}\right),\label{eq:SnBM}
\end{align}
where $\tilde{\rho}_m := P_m \rho P_m$ is the unnormalized density matrix after measurements.
These replicated quantities are analytically tractable and capture the qualitative behaviors of $S_{B|M}$ despite exhibiting a different universality at the critical point~\cite{bao2020theory,jian2020measurement}.
To study the transition in $S_{B|M}$, one needs to consider $S^{(n)}_{B|M}$ for all $n$ and then take the replica limit $n \to 1$.

To gain qualitative insights, we focus on $S^{(2)}_{B|M}$, which can be studied analytically via a mapping to the effective Hamiltonian. 
The essence of the mapping is to identify the average second moments $\overline{\tr\tilde{\rho}_{B,m}^2}$ and $\overline{\tr\tilde{\rho}_{m}^2}$ in Eq.~\eqref{eq:SnBM} with the partition function of a classical spin model with certain boundary conditions.
To establish the mapping, we formulate the dynamics of the average purity, which involves two copies of density matrix, as the evolution of state vector $\kket{\rho} \simeq \rho \otimes \rho$ in the duplicated Hilbert space $\mathcal{H}^{(2)} = (\mathcal{H} \otimes \mathcal{H}^*)^{\otimes 2}$.
Computing the subsystem purity is given by the overlap between $\kket{\rho}$ and a reference state, which will manifest as boundary conditions in the classical spin model at final times~\cite{bao2021symmetry,block2022measurement}.

To start with, the dynamics of $\kket{\rho}$ in the random unitary circuits [illustrated in Fig.~1(a) of the main text] is generated by an unitary operator $\mathcal{U} = (U \otimes U^*)^{\otimes 2}$ in $\mathcal{H}^{(2)}$.
Two copies of $U$ and $U^*$ in $\mathcal{U}$ act on ket and bra vector spaces, respectively.
The output state can be expressed as $\kket{\rho(t)} = \prod_{\tau = 1}^{N_t} \mathcal{U}_{2,\tau}\mathcal{U}_{1,\tau} \kket{\rho_0}$,
where $\mathcal{U}_{1,\tau}$ and $\mathcal{U}_{2,\tau}$ are the single-qubit and two-qubit Haar random gates in the $\tau$-th time step, respectively, and $N_t$ is the total number of time step, i.e. $t = N_t \delta t$.

Averaging over the single-qubit unitary gates yields a projector onto a two-dimensional reduced local Hilbert space
\begin{align}
    \overline{(U_{j,\tau}\otimes U_{j,\tau}^*)^{\otimes 2}} = \frac{1}{3}\kket{\mathcal{I}_j}\bbra{\mathcal{I}_j} + \frac{1}{3} \kket{\mathcal{C}_j}\bbra{\mathcal{C}_j} - \frac{1}{6} \kket{\mathcal{I}_j}\bbra{\mathcal{C}_j} - \frac{1}{6} \kket{\mathcal{C}_j} \bbra{\mathcal{I}_j},
\end{align}
where $\kket{\mathcal{I}_j} \equiv \sum_{ab} \kket{aabb}$ and $\kket{\mathcal{C}_j} \equiv \sum_{ab} \kket{abba}$ with $a,b$ run over the local Hilbert space of qubit $j$.
The coefficients on the right-hand side are given by the Weingarten function for a single-qubit random unitary~\cite{collins2003moments,nahum2018operator}.
We note that $\kket{\mathcal{I}}$ and $\kket{\mathcal{C}}$ are not orthogonal and choose an orthonormal basis labeled by a spin-$1/2$ variable $s_{j,\tau} = \uparrow, \downarrow$ such that
\begin{align}
    \overline{(U_{j,\tau}\otimes U_{j,\tau}^*)^{\otimes 2}} = \sum_{s_{j,\tau} = \uparrow,\downarrow} \kket{s_{j,\tau}}\bbra{s_{j,\tau}}.
\end{align}
We choose to define Pauli matrices such that the eigenstates of $\sigma^x_j$ are given by
\begin{align}
    \kket{+} &= \frac{1}{\sqrt{2}} \left(\kket{\uparrow} + \kket{\downarrow}\right) \equiv \frac{1}{2\sqrt{3}}(\kket{\mathcal{I}_j} + \kket{\mathcal{C}_j}), \\
    \kket{-} &= \frac{1}{\sqrt{2}} \left(\kket{\uparrow} - \kket{\downarrow}\right) \equiv \frac{1}{2}(\kket{\mathcal{I}_j} - \kket{\mathcal{C}_j}).
\end{align}

Analogously, the average two-qubit unitary operation $\overline{\mathcal{U}_{ij,\tau}}$ on site $i$ and $j$ leads to a projector onto reduced Hilbert space of two spins
\begin{align}
    \overline{\mathcal{U}_{ij,\tau}} = \overline{(U_{ij,\tau} \otimes U_{ij,\tau}^*)^{\otimes 2}} = \frac{1}{15}\kket{\mathcal{I}_i\mathcal{I}_j}\bbra{\mathcal{I}_i\mathcal{I}_j} + \frac{1}{15} \kket{\mathcal{C}_i\mathcal{C}_j}\bbra{\mathcal{C}_i\mathcal{C}_j} - \frac{1}{60} \kket{\mathcal{I}_i\mathcal{I}_j}\bbra{\mathcal{C}_i\mathcal{C}_j} - \frac{1}{60} \kket{\mathcal{C}_i\mathcal{C}_j} \bbra{\mathcal{I}_i\mathcal{I}_j},
\end{align}
where the coefficients on the right-hand side are given by the Weingarten functions for a random unitary on two qubits~\cite{collins2003moments,nahum2018operator}.

In each time step $\delta t$, the layer of two-qubit gates $\mathcal{U}_{2,\tau}$ consists of $N\delta t$ gates $\mathcal{U}_{ij,\tau}$.
Each gate acts on a pair of qubits at site $i$ and $j$ drawn randomly from distribution $P(i,j)$. 
The projection of $\overline{\mathcal{U}_{2,\tau}}$ onto the reduced Hilbert space of each site defines a transfer matrix $\mathcal{T} \equiv \bbra{\{s_{j,\tau+1}\}} \overline{\mathcal{U}_{2,\tau}} \kket{\{s_{j,\tau}\}}$ that describes the transition amplitude between the effective Hilbert space at two consecutive time steps.
In the limit $\delta t \to 0$, the transfer matrix can be written as the imaginary time evolution $\mathcal{T} = e^{-\delta t H_{\text{eff}}}$ generated by the effective quantum Hamiltonian of the form [Eq.~(1) in the main text]
\begin{align}
    H_{\text{eff}} &= \sum_{i,j}J_{ij} \left[ -\frac{2}{5}\sigma_i^z\sigma_j^z - \frac{1}{10}\sigma_i^y \sigma_j^y - \frac{1}{5}(\sigma_i^x + \sigma_j^x) \right].
\end{align}
We note that the effective Hamiltonian $H_{\text{eff}}$ exhibits an Ising symmetry generated by $\prod_{i}\sigma^x_i$, which originates from the invariance of $\mathcal{U}$ under swapping two copies of $U$.

Having established the mapping for the dynamics of double density matrix, we now discuss the boundary conditions associated with $\overline{\tr \tilde{\rho}_{B,m}^2}$.
In the output state, we first perform measurements on $M$, which enforce a projection on the replicated density matrix by $\mathcal{P}_m = P_m^{\otimes 4}$.
Then, we compute the purity of subsystem $B$ which is given by the overlap 
\begin{align}
    \tr\tilde{\rho}_{B,m}^2 = \bbra{\mathcal{I}}\mathcal{C}_B \mathcal{P}_m\kket{\rho(t)},\label{eq:purity}
\end{align}
where $\mathcal{C}_B$ is the swap operator of two copies of ket vectors in region $B$, and $\mathcal{C}_B\kket{\mathcal{I}} = \kket{\mathcal{C}_B}$.

Measurements enforce symmetric boundary conditions on $M$ at final times since the reference state $\bbra{\mathcal{I}}\mathcal{P}_{m} = \bbra{mmmm}$ is symmetric under swapping of two ket vectors.
Similarly, the initial product state also enforces open boundary conditions as $\kket{\rho_0} = \kket{0000}$.
At the position of the input qubit $A$ and output qubit $B$, Eq.~\eqref{eq:purity} imposes symmetry breaking boundary conditions: $\bbra{\mathcal{I}_A}$ at site $i_A$ and $\bbra{\mathcal{C}_B}$ at site $i_B$.
The overlap contributes a Boltzmann weight $e^{-hs}$
\begin{align}
    \bbrakket{\mathcal{I}_A}{s_{A}} &= \frac{\sqrt{3}}{2} + \frac{s_{A}}{2}, \\
    \bbrakket{\mathcal{C}_B}{s_{B}} &= \frac{\sqrt{3}}{2} - \frac{s_{B}}{2},
\end{align}
which corresponds to a magnetic field $h = \log(\sqrt{3/2}-\sqrt{1/2}) = - 0.66$ for $\mathcal{I}$ boundary conditions and $-h$ for $\mathcal{C}$ boundary conditions.
We use $s_A$ and $s_B$ to label the basis states of the Hilbert space at spacetime locations $(i_A, 1)$ and $(i_B, N_t)$, respectively.
For the denominator $\overline{\tr\tilde{\rho}_m^2}$ in Eq.~\eqref{eq:SnBM}, $\mathcal{I}$ boundary conditions are imposed at both $A$ and $B$.

Hence, $S^{(2)}_{B|M}$ is given by
\begin{align}
    S^{(2)}_{B|M} = -\log\left(\frac{\bbra{\mathcal{I}}\mathcal{P}_m e^{ h \sigma^z_{B}} e^{-t H_{\text{eff}}} e^{-h \sigma^z_{A}}\kket{\rho_0}}{\bbra{\mathcal{I}}\mathcal{P}_m e^{ - h \sigma^z_{B}} e^{-t H_{\text{eff}}} e^{-h \sigma^z_{A}} \kket{\rho_0}}\right).
\end{align}
where $\sigma_{A/B}^z$ is the Pauli-Z operator at site $A/B$.
In the case that magnetization is small, e.g. close to the critical point, one can treat the magnetic field perturbatively.
We expand $S^{(2)}_{B|M}$ to second order in $h$ and obtain 
\begin{align}
    S^{(2)}_{B|M} = 2h^2  \frac{\bbra{\mathcal{I}}\mathcal{P}_m \sigma^z_{B} e^{-t H_{\text{eff}}} \sigma^z_{A}\kket{\rho_0}}{\bbra{\mathcal{I}}\mathcal{P}_m e^{-t H_{\text{eff}}} \kket{\rho_0}} := 2h^2 \langle \sigma^z_B(t) \sigma^z_A(0)\rangle.
\end{align}
Hence, $S^{(2)}_{B|M}$ is proportional to the imaginary time order parameter correlation function $\langle \sigma^z_B(t) \sigma^z_A(0)\rangle$ near the critical point and reflects the universal properties of the phase transition in the effective spin model.

\section{Mean-field theory for finite-temperature transition in all-to-all coupled quantum Ising model}

\begin{figure}
    \centering
    \includegraphics[width=0.7\textwidth]{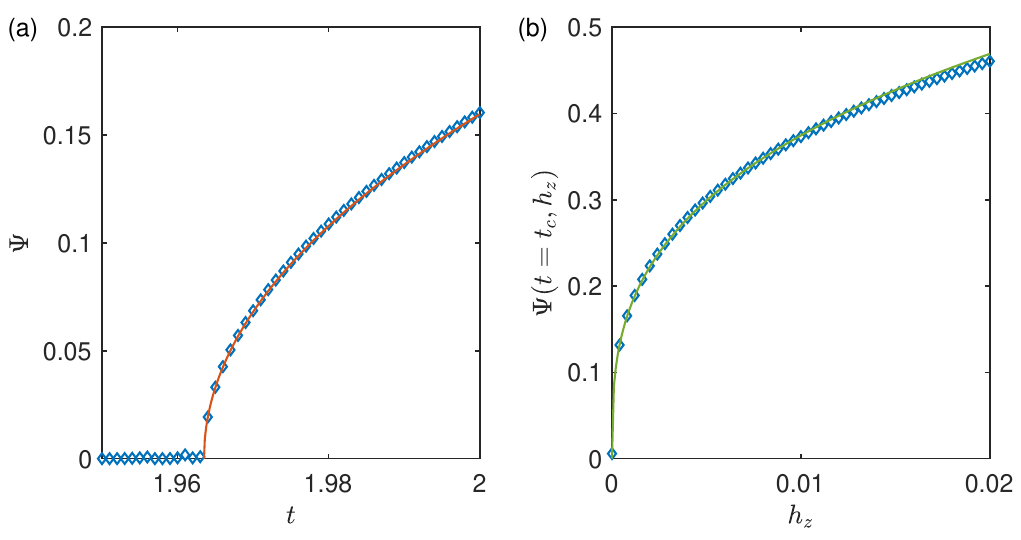}
    \caption{Mean-field theory of the ordering phase transition in the effective classical spin model of the all-to-all coupled quantum circuit. (a) Average magnetization $\Psi$ as a function of time $t$. We obtain critical time $t_c = 1.96$ and critical exponent $\beta = 0.49 \pm 0.04$. The red solid line represents $\Psi = c_1(t - t_c)^\beta$ for $t > t_c$. 
    (b) Critical scaling of magnetization $\Psi$ as a function symmetry breaking field $h_z$. We obtain $h_z \sim \Psi^\delta$ with $\delta = 3.1\pm 0.1$ at the critical point. The green solid line represents $\Psi(t_c, h_z) = c_2h_z^{1/\delta}$.
    }
    \label{suppfig:MF_all_to_all}
\end{figure}

In this section, we perform the exact mean-field calculation for the phase transition in the effective Hamiltonian for all-to-all coupled random unitary circuits.
We first perform the quantum-to-classical mapping. 
Then, we use mean-field theory, which is exact in the thermodynamic limit, to determine the critical time and critical exponents.

The effective quantum Hamiltonian of the all-to-all coupled RUC is given by
\begin{align}
    H_{\text{eff}} = \sum_{(i,j)} h_{ij} = \sum_{(i,j)} \frac{2}{N-1} \left[-\frac{2}{5}\sigma_i^z \sigma_j^z - \frac{1}{10}\sigma_i^y \sigma_j^y - \frac{1}{5}(\sigma_i^x + \sigma_j^x) \right],
\end{align}
where $(i,j)$ represents a pair of qubits on sites $i$ and $j$.
We note that $J_{ij} = 2/(N-1)$ such that $\sum_{(i,j)} J_{ij} = N$.
The partition function of the effective spin model for the circuit of finite time $t$ is
\begin{align}
    Z &= \int Ds \; e^{-\int_{0}^t \rd \tau H_{\text{eff}}}.
\end{align}
The only difference between this partition function and that of the quantum Hamiltonian at finite temperature are the boundary conditions in the temporal direction.
The spin model we derived takes open boundary conditions at $\tau = 0$ and $t$.
If one assumes periodic temporal boundary conditions, the finite temperature transition can be analyzed by standard mean-field theory, which gives rise to critical exponents $\beta = 1/2$ and $\nu = 2$.

To incorporate the effects of open temporal boundary conditions, we perform the quantum-to-classical mapping.
First, we divide the time interval $[0, t]$ into $N_t$ steps, i.e. $t = N_t \delta t$, and index these time steps by $\tau$.
Within each time step $\tau$, we Trotterize the imaginary time evolution into the product of $N(N-1)/2$ terms as $e^{-\delta tH_{\text{eff}}} = \prod_{(i,j)}e^{-\delta t h_{ij}}$.
Then, we insert resolutions of the identity using the eigenstates of $\sigma_i^z$, labeled by a classical spin $s_i = \pm1$, before and after each Trotterized time step $\zeta$. 
The classical spins at two consecutive Trotterized time steps $\zeta$ and $\zeta+1$ are coupled by the transfer matrix
\begin{align}
    \mathcal{T}_{ij,\tau} \equiv& \left< s_{i,\tau,\zeta+1} s_{j,\tau,\zeta+1} \right| e^{-\delta t h_{ij} } \left| s_{i,\tau,\zeta} s_{j,\tau,\zeta}\right> \nonumber \\
    =& -\log(1-2h)\frac{1+s_{i,\tau,\zeta+1} s_{i,\tau,\zeta}}{2} \frac{1 + s_{j,\tau,\zeta+1}s_{j,\tau,\zeta}}{2} - \log(h)\frac{1 - s_{i,\tau,\zeta+1}s_{i,\tau,\zeta}s_{j,\tau,\zeta+1}s_{j,\tau,\zeta}}{2} \\
    &- \left( \log(J_{yy})+ \ri\pi\frac{1 + s_{i,\tau,\zeta} s_{j,\tau,\zeta}}{2}\right) \frac{1-s_{i,\tau,\zeta+1} s_{i,\tau,\zeta}}{2} \frac{1 - s_{j,\tau,\zeta+1}s_{j,\tau,\zeta}}{2} - J_{zz} s_{i,\tau,\zeta} s_{j,\tau,\zeta},\nonumber 
\end{align}
where $h = 2\delta t/5(N-1)$, $J_{yy} = \delta t/5(N-1)$, and $J_{zz} = 4\delta t/5(N-1)$.

In terms of the classical spins $s_{i,\tau,\zeta}$, the partition function is
\begin{align}
    Z = \sum_{\{s_{i,\tau,\zeta}\}} \prod_{\tau = 1}^{N_t} \prod_{(i,j)} \mathcal{T}_{ij,\tau}.
\end{align}
In the following, we analyze the phase transition as a function of $t$.

We use mean-field theory, which is exact in the thermodynamic limit $N \to \infty$. 
To this end, we introduce the mean fields $\Psi_{\tau,\zeta} = \sum_i \langle s_{i,\tau,\zeta} \rangle/N$ and $\Phi_{\tau,\zeta} = \sum_i \langle s_{i,\tau,\zeta}s_{i,\tau,\zeta+1}\rangle /N$. 
This allows one to factor the transfer matrix as $\mathcal{T}_{ij,\tau} = \mathcal{T}^{(MF)}_{i,\tau} \mathcal{T}^{(MF)}_{j,\tau}$, yielding
\begin{align}
    Z = \prod_i Z_{1d,i} = \prod_i \left[\sum_{\{s_{i,\tau,\zeta}\}} \prod_{\tau = 1}^{N_t} \left( \mathcal{T}^{(MF)}_{i,\tau} \right)^{N-1}\right],
\end{align}
where the partition function $Z$ decouples into $N$ $1$d classical Ising models in the temporal direction for spins living on every spatial site, $Z_{1d,i} = \sum_{\{s_{\tau,\zeta}\}}e^{-H_{1d}}$.
We note that, after the Trotterization, $N-1$ transfer matrices $\mathcal{T}_{ij,\tau}$ act on site $i$.

Before we derive the mean-field solution, we can first simplify the classical model.
Each Trotterized time step $e^{\delta t h_{ij}}$ can only  change the order parameter by at-most $O(\delta t/N)$, where the factor $1/N$ from the coupling $J_{ij} \sim 1/N$.
Hence, $\Phi_{\tau,\zeta}$ which measures the correlation in two consecutive Trotterized steps is given by $\Phi_{\tau,\zeta} = 1 + O(\delta t/N)$.
We can therefore, to the leading order, replace $\Phi_{\tau,\zeta}$ by unity.
Moreover, we drop $\zeta$ dependence in $\Psi_{\tau,\zeta}$ because $\Psi_{\tau,\zeta}$ is slowly varying within each time step, i.e. we approximate $\Psi_{\tau,\zeta} \approx \Psi_{\tau}$ in $\delta t$, where $\Psi_{\tau}$ is the average $\Psi_{\tau,\zeta}$ over Trotterized steps $\zeta$.

After the simplification, the $1$d classical Hamiltonian takes the form
\begin{align}
    H_{1d} = \sum_{\tau = 1}^{N_t}\sum_{\kappa = 1}^{N-1} -\tilde{J}_\tau s_{\tau,\kappa}s_{\tau,\kappa+1} - \tilde{h}_\tau s_{\tau,\kappa},\label{eq:H1d}
\end{align}
where $s_{\tau,\kappa}$ are spins inserted before and after $N-1$ transfer matrices $\mathcal{T}^{(MF)}_{i,\tau}$ labeled by $\kappa$, and $s_{\tau,N} \equiv s_{\tau+1,1}$.
The couplings in the $1$d Ising model to the leading order in $1/N$ are given by
\begin{align}
    \tilde{J}_\tau &= - \frac{\log h}{2},\\
    \tilde{h}_\tau &= J_{zz}\Psi_\tau.
\end{align}

We determine the critical point and critical exponents by numerically solving the self-consistency equations
\begin{align}
    \Psi_\tau = \frac{1}{Z_{1d}} \sum_{\{s_\tau\}} s_\tau e^{-H_{1d}}.\label{eq:sce}
\end{align}
The global magnetization $\Psi$ as a function of time is plotted in Fig.~\ref{suppfig:MF_all_to_all}(a).
We obtain a critical time $t_c = 1.96$. 
Near the critical time, $\Psi \sim (t - t_c)^\beta$ with the order parameter critical exponent $\beta = 0.49 \pm 0.04$.
We determine the other critical exponent by considering the scaling of the order parameter as a function of external magnetic field $h_z$ at the critical time, i.e. $h_z \sim \Psi^\delta$.
The external magnetic field introduces an additional term to the $1$d Hamiltonian [Eq.~\eqref{eq:H1d}]
\begin{align}
    H_m(h_z) = \sum_{\tau = 1}^{N_t} \sum_{\kappa = 1}^{N-1} -\frac{h_z\delta t}{N-1} s_{\tau,\kappa}.
\end{align}
Now solving the self-consistency equations in the presence of $h_z$ yields $\delta = 3.1 \pm 0.1$ as shown in Fig.~\ref{suppfig:MF_all_to_all}(b).
Using the scaling relations in $1$d, we have $\nu = \beta(\delta + 1) = 2.0 \pm 0.2$.
The numerically extracted critical exponents agree with the standard mean-field exponents $\beta = 1/2$, $\delta = 3$, and $\nu = 2$.

\section{Details of finite-time transition in 1d long-range circuits}
The effective quantum Hamiltonian predicts ordering phase transitions for $\alpha < 2$~\cite{ruelle1968statistical,dyson1969existence,thouless1969long,anderson1971some}, a Kosterlitz-Thouless (KT) like phase transition at $\alpha = 2$~\cite{kosterlitz1976phase,cardy1981one,bhattacharjee1981some,bhattacharjee1982n,imbrie1988intermediate, luijten2001criticality}, and the absence of a phase transition for $\alpha > 2$.
In this section, we present details concerning our numerical evidence for these qualitative predictions, as well as on our estimation of the critical exponents shown in Fig.~2(c) of the main text.

\subsection{Phase transitions for \texorpdfstring{$\alpha < 2$}{alpha<2}}
\begin{figure}
    \centering
    \includegraphics[width=0.9\textwidth]{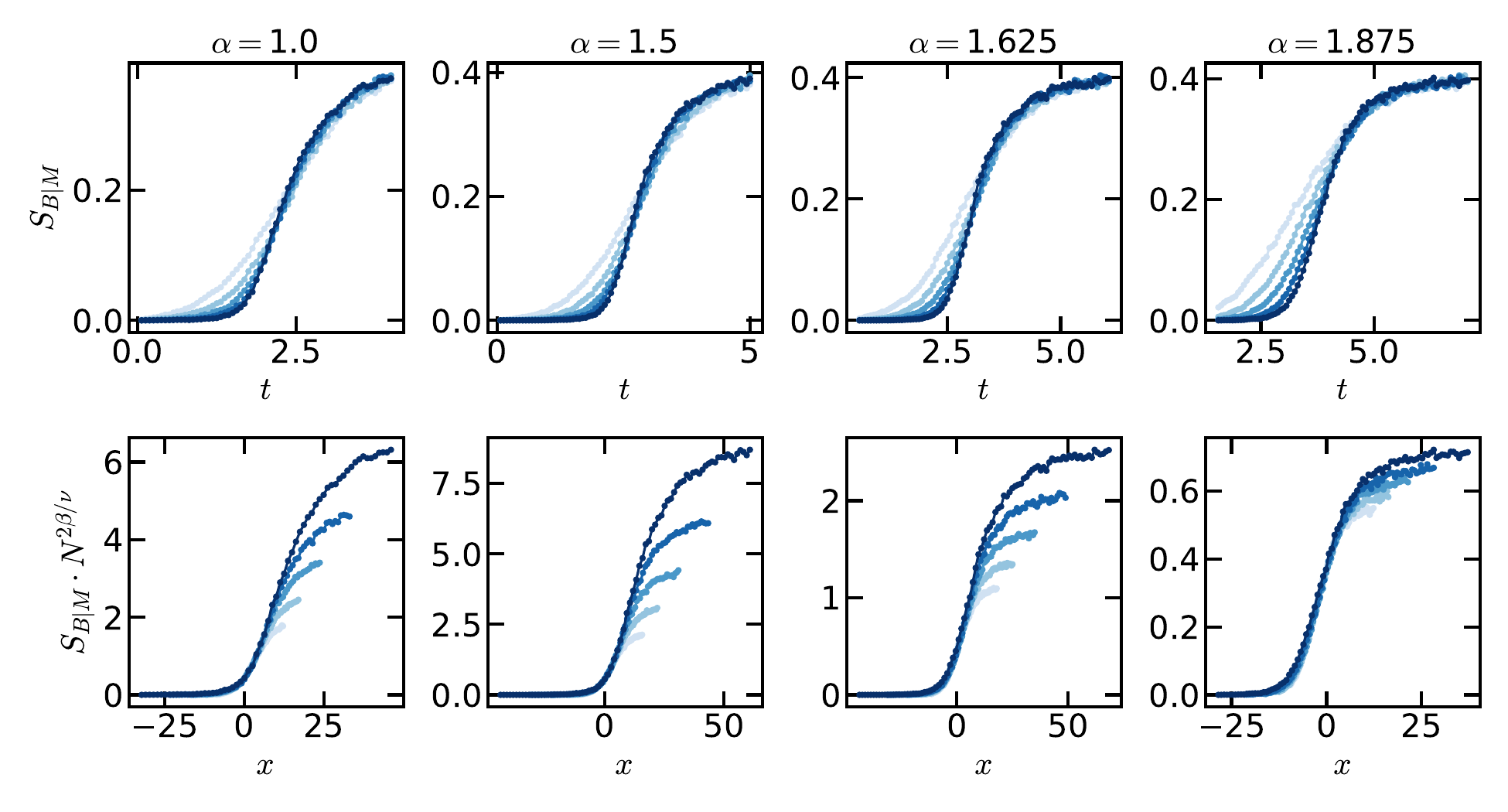}
    \caption{(Top row) Finite time teleportation transition in $1$d long-range circuits with $\alpha = 1.0, 1.5, 1.625, 1.875$ (from left to right) for various system sizes $N$ from $32$ to $512$ indicated by increasing opacity. (Bottom row) Finite-size scaling collapse using Eq.~(3) in the main text. The numerically extracted critical exponents are presented in Fig.~2(c) in the main text. The $x$-axis represents $(t - t_c)N^{1/\nu}$. The results were averaged over $1.5 \cdot 10^4$ circuit realizations (as for $\alpha=0.0, 1.75$ in the main text).
    }
    \label{suppfig:transitions}
\end{figure}

Based on the mapping of the conditional entropy to the order parameter correlation function in the effective spin model, we can derive a finite-size scaling formula.
Assuming the second-order phase transition, the order parameter $\langle \sigma^z \rangle$ would vanish as $\langle \sigma^z \rangle \sim (t - t_c)^\beta$ close to the critical point in the ordered phase ($t > t_c$).
Here, we use $\beta$ to denote the order parameter critical exponent.
Accordingly, the conditional entropy decays as $S_{B|M} \sim \langle \sigma^z_B(t) \sigma^z_A(0)\rangle \sim (t-t_c)^{2\beta}$.
In numerical simulation, the singularity at $t = t_c$ is smeared out due to the finite size effect. 
Specifically, we have $S_{B|M} = (t-t_c)^{2\beta}f(N/\xi)$, where $f(N/\xi)$ is a universal function that only depends on the ratio between the system size $N$ and the correlation length $\xi$.
Knowing that the correlation length diverges as $\xi \sim (t-t_c)^{-\nu}$, we can obtain the scaling formula in Eq.~(3) of the main text, which we recast here
\begin{align}
    S_{B|M} = N^{-2\beta/\nu}\mathcal{F}((t - t_c)N^{1/\nu}).
\end{align}

Figure~\ref{suppfig:transitions} shows the numerical results for $S_{B|M}(t,N)$ (top row) and associated finite-size scaling (FSS) collapse (bottom row) for $\alpha = 1.0, 1.5, 1.625, 1.875$ (analogous plots for $\alpha=0.0, 1.75$ are in the main text).
The analysis yields the critical exponents $\nu$ and $\beta$, and the critical time $t_c$.
Specifically, we optimize the least-squared error (LSE) of the fit of $S_{B|M}(t,N)$ as a function of $\beta$, $\nu$, and $t_c$.
We weight errors in the collapse according to a Gaussian distribution centered near the critical time with standard deviation $40$ for the parameter $(t-t_c)L^{1/\nu}$.
The critical exponents are close to their mean-field values below $\alpha = 1.5$ and vary continuously when $1.5 < \alpha < 2$, which is consistent with the prediction from the effective Hamiltonian.
We note that the effective Hamiltonian also predicts the divergence of $\nu$ as $\nu = 1/(2 -\alpha)$~\cite{kosterlitz1976phase}.
To confidently examine this behavior requires a more accurate scheme to extract $\nu$, which is left for future work.

To obtain estimates and errors for our FSS analysis, we employ a standard bootstrap scheme: (1) $15000$ circuit realizations are simulated for $N \in \{32, 64, 128, 256, 512\}$, from which we obtain samples of $S_{B|M}(t,N)$; (2) we select $7500$ random sub-samples which we average to estimate $S_{B|M}(t,N)$; (3) we perform 3-parameter curve-fitting to extract samples $\nu$, $\beta$ and $t_c$; (4) steps (2-3) are repeated 1000 times to obtain distributions for $\nu$, $\beta$ and $t_c$.
We report the mean of the distribution as our estimate and all error bars reflect one standard deviation from the mean. 
This procedure is used for all FSS analysis \emph{except} $\alpha=2.0$, which requires special attention due to the expected failure of scaling form Eq.~(3) (see discussion below).

We note that the saturation value of $S_{B|M}$ at a long time to $0.4$ is universal for the Clifford simulation.
After projective measurements on $M$, the unmeasured qubit $A$ and $B$ are in the \emph{projective ensemble} defined by measurement results and circuits realizations~\cite{cotler2023emergent,choi2023preparing}.
In the projective ensemble of deep Clifford circuits, $A$ and $B$ are entangled by a two-qubit random Clifford gate with an average entropy $0.4$, which can be shown analytically.

\subsection{Kosterlitz-Thouless like transition at \texorpdfstring{$\alpha = 2$}{alpha=2}}
\begin{figure}[t!]
    \centering
    \includegraphics[width=0.98\textwidth]{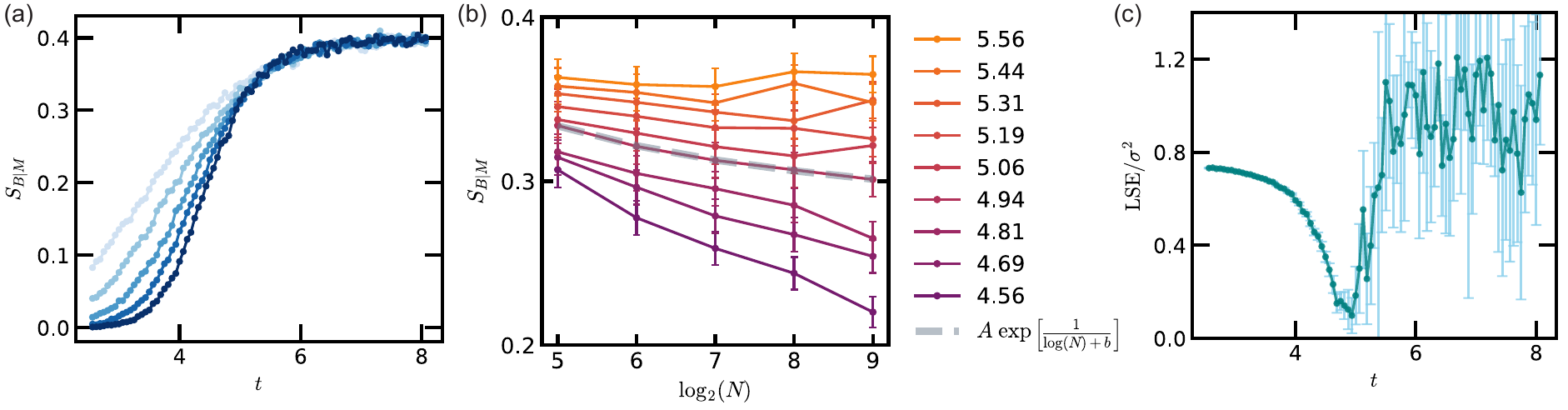}
    \caption{Finite-time teleportation transition in $1$d long-range circuits with $\alpha = 2$. (a) Conditional entropy $S_{B|M}$ as a function of time $t$ for various system sizes $N$ from $32$ to $512$ indicated by increasing opacity.
    The results are averaged over $15000$ random circuit realizations.
    (b) $S_{B|M}$ as a function of $N$ for various time $t$ given in the legend.
    Dash-dotted line indicates the best fit of critical scaling form $S_{B|M}(t_c, N) = A\exp[1/(\log N+b)]$ for a KT-like transition.
    (c) LSE (obtained fitting $S_{B|M}$ to the critical scaling form $A\exp[1/(\log N + b)]$) divided by the variance $\sigma^2$. 
    Error bars on this figure of merit (FOM) are obtained from 1000 ``synthetic" simulations where noise is added to $S_{B|M}(t, N)$ according to the observed noise in the true simulation.
    The critical time $t_c = 4.95 \pm 0.12$ is determined by the minimum of the FOM (error obtained from the same synthetic simulations).
    }
    \label{suppfig:KT}
\end{figure}

The effective quantum Hamiltonian for the $1$d long-range random circuit at $\alpha = 2$ is given by
\begin{align}
    H_{\text{eff}} = \sum_{i,j} \frac{J}{|i-j|^2} \left[-\frac{2}{5}\sigma_i^z \sigma_j^z - \frac{1}{10}\sigma_i^y \sigma_j^y - \frac{1}{5}(\sigma_i^x + \sigma_j^x) \right].
\end{align}
At finite temperature, the model is in the same universality as the $1$d classical Ising model with inverse square interaction, which exhibits a finite-temperature Kosterlitz-Thouless phase transition~\cite{kosterlitz1976phase,cardy1981one,bhattacharjee1981some,bhattacharjee1982n,imbrie1988intermediate, luijten2001criticality}.

According to the mapping discussed above, the conditional entropy is related to the order parameter correlation function, i.e. $S^{(2)}_{B|M} \simeq \langle \sigma^z_B(t) \sigma^z_{A}(0) \rangle$.
Near the critical point, the correlation function $\langle \sigma^z_B(t) \sigma^z_{A}(0) \rangle$ is of the same asymptotic form as the correlation $G(r) := \langle s(r) s(0)\rangle$ in the 1d classical Ising model, where $r$ is the distance between qubits $A$ and $B$.
Using the renormalization group method developed in Ref.~\cite{yuval1970exact}, the scaling form of $G(r)$ has been derived~\cite{bhattacharjee1981some}. 
In the ordered phase close to the critical point, $G(r)$ exhibits a subleading power-law decay
\begin{align}
    G(r) = \bar{s}^2 \left(1 + \frac{4\sqrt{|(T - T_c)/T_c|}}{r^{4\sqrt{|(T - T_c)/T_c|}}}\right), \quad (T < T_c).
\end{align}
At the critical point, the scaling changes to
\begin{align}
    G(r) = \bar{s}^2 e^{1/\ln r}, \quad (T = T_c).
\end{align}

In the numerical simulation, we obtain $S_{B|M}$ as a function of $t$ for various system sizes $N$.
We fix the distance between $A$ and $B$ to be $N/2$.
Hence, at the critical point, $S_{B|M}$ exhibits the finite-size scaling
\begin{align}
    S_{B|M}(t_c, N) = A\exp\left(\frac{1}{\ln N + b}\right).
\end{align}
To determine the critical point, we fit $S_{B|M}(t,N)$ as a function of $N$ for each $t$ to the critical scaling form.
The LSE divided by the total variance is plotted in Fig.~\ref{suppfig:KT}(c) as a function of time.
We determine the critical time $t_c = 4.95 \pm 0.12$ according to the minimum of the LSE/variance curve -- this is the time at which the most variance in the raw data is explained by the critical scaling function.
The numerical data at $t_c$ seems to agree with the predicted critical scaling function, as shown in Fig.~\ref{suppfig:KT}(b), where the dashed-dotted line denotes the expected critical behavior.
We note that although the KT scaling is predicted from an approximate quantity $S_{B|M}^{(2)}$ and may not hold in the replica limit $n \to 1$, a KT-like transition is predicted for any number of replicas with $\alpha = 2$ long-range interaction~\cite{cardy1981one,luijten2001criticality}.
We thus expect a KT-like universality to persist in the replica limit $n \to 1$.
We leave it for a future study to conclusively determine this universality.

\subsection{Absence of phase transition for \texorpdfstring{$\alpha > 2$}{alpha>2}}
\begin{figure}[t!]
    \centering
    \includegraphics[width=0.35\textwidth]{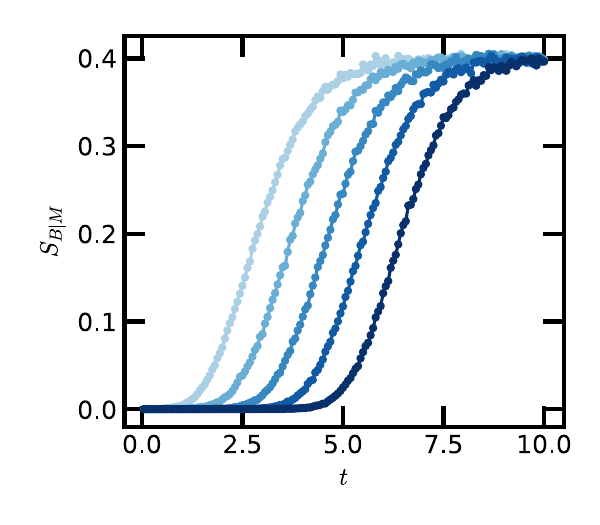}
    \caption{Absence of finite-time teleportation transition in $1$d long-range circuits with $\alpha = 3.0$. $S_{B|M}$ is plotted as a function of $t$ for various system sizes $N$ from $32$ to $512$. The numerical results are averaged over $3.0 \cdot 10^4$ random circuit realizations.}
    \label{suppfig:no_transition}
\end{figure}

The effective Hamiltonian does not exhibit finite temperature phase transition for $\alpha > 2$. As an example, in the circuit with $\alpha = 3$, we compute the conditional entropy as a function of $t$ for various system sizes (see Fig.~\ref{suppfig:no_transition}). 
Notably, the $S_{B|M}$ curves do not exhibit a crossing with varying system sizes, indicating the lack of a finite-time singularity in the thermodynamic limit.

\bibliography{refs}